\documentclass[prb,preprint]{revtex4-1} 

\usepackage{amsmath}
\usepackage{amsfonts}
\usepackage{graphicx}

\newcommand{\ev}[1]{\left\langle #1 \right\rangle}
\newcommand{\half}{\tfrac{1}{2}}

\begin{document}


\title{Is the mean free path the mean of a distribution?}

\author{Steve T. Paik}
\email{paik\_steve@smc.edu}
\affiliation{Physical Science Department, Santa Monica College,
Santa Monica, CA 90405}

\date{\today}

\begin{abstract}
We bring attention to the fact that Maxwell's mean free path for a
dilute hard-sphere gas in thermal equilibrium, $(\sqrt{2}\sigma n)^{-1}$, 
which is ordinarily obtained by multiplying the average speed by the average
time between collisions, is also the statistical mean of the distribution
of free path lengths in such a gas.
\end{abstract}

\maketitle

\section{Introduction}

For a gas composed of rigid spheres (``molecules'')
each molecule affects another's motion only at collisions ensuring
that each molecule travels in a straight line at constant speed between
collisions --- this is a free path. The mean free path, as the name suggests,
should then be the average length of a great many free paths either made by an
ensemble of molecules or a single molecule followed for a long time, it being a
basic assumption of statistical mechanics that these two types of averages are
equivalent. However, the mean free path is ordinarily defined in textbooks as
the ratio of the average speed to the average frequency of collisions. Are
these two different ways of defining the mean free path equivalent? In this
article we answer this question in the affirmative for a hard-sphere gas and,
along the way, discuss aspects of the classical kinetic theory that may be of
interest in an upper-level undergraduate course on thermal physics.

Consider a gas of hard spherical particles of mass $m$ and radius
$a$ which have attained thermal equilibrium and spatial uniformity.
The mean free path quoted in textbooks is $(\sqrt{2}\sigma n)^{-1}$, where
$\sigma$ is the total scattering cross section and $n$ is the average number
density throughout the gas.\cite{Reif}
In three dimensions, $\sigma = \pi(2a)^2$.
This expression for the mean free path is valid only in the limit where the
gas is dilute, 
but not so dilute that the mean free path becomes comparable to the width
of the container.
In this article we will discuss a slightly more general form of the mean free
path: $(\sqrt{2}\sigma n\chi(n))^{-1}$, which may be applied even when the
gas is not dilute. Here $\chi(n)$ may be regarded as a
finite-density correction that starts from a value of one in the extreme
dilute limit and increases monotonically with density.\cite{Chapman}
When the density of the spheres approaches that value obtained by packing the
spheres as closely as possible, $\chi^{-1}$ vanishes so that the mean free path
is zero.\cite{cp} The factor $\chi$ has been referred to in the literature
as Enskog's $\chi$ due to its appearance in the Enskog kinetic theory of
transport.

The expression with $\chi = 1$ may be attributed to Maxwell who defined the mean
free path $\lambda_\textrm{Maxwell}$ as the ratio of
the mean speed $\ev{v}$, evaluated in the equilibrium velocity distribution
$f(\mathbf{v})$, to the average collision frequency per molecule $\omega$:
\begin{equation}
\label{Maxwellmfp}
  \lambda_\textrm{Maxwell} \equiv \frac{\ev{v}}{\omega}.
\end{equation}
The average collision frequency is the reciprocal of the average time between
collisions so Eq.~(\ref{Maxwellmfp}) is simply the product of the average
molecular speed and the average time of flight between collisions.
Working in scaled units where masses are measured in units of $m$ and
velocities are measured in units of $(kT/m)^{1/2}$, the velocity probability
density is $f(\mathbf{v}) = (2\pi)^{-3/2} \exp(-v^2/2)$.
$T$ is the temperature and $k$ is Boltzmann's
constant. Angled brackets denote an instruction to take an
average in the equilibrium velocity distribution:
$\ev{\dotsc} \equiv \int \dotsc f(\mathbf{v})d\mathbf{v}$. 

The average collision frequency $\omega$ may be obtained through a standard
flux argument wherein one counts the total number of collisions
$\omega\Delta t$ received by a randomly chosen target molecule 
during an arbitrary time interval $\Delta t$. Let us label the
target molecule ``1;''
there is a probability $f(\mathbf{v}_1)d\mathbf{v}_1$ to find
it with velocity centered around some $\mathbf{v}_1$.
Along any particular direction through 1 there will be an incoming
stream of scatterers all moving, roughly speaking, in the general direction
toward 1. For a subset of these scatterers with velocities centered around 
$\mathbf{v}_2$, only those in a cylindrical volume
$|\mathbf{v}_1-\mathbf{v}_2|\sigma\Delta t$
can reach the target in the allotted time. These scatterers do
not necessarily have collinear flight paths, some will have oblique paths that
might cause them to exit through the side wall of the imaginary cylinder prior
to making contact with 1. Therefore, one imagines making $\Delta t$ sufficiently
small to ensure that all molecules in the cylinder do collide with 1. 
The number of scatterers in this cylinder is obtained by multiplying the
volume by a density yielding
$|\mathbf{v}_1-\mathbf{v}_2|\sigma\Delta t\, n f(\mathbf{v}_2)d\mathbf{v}_2$.
The condition of molecular chaos is assumed: in the
encounter between two spheres their initial velocities are uncorrelated prior
to the collision. This assumption allows one to express the collision
probability in terms of a product of equilibrium velocity distributions.
Integrating over all possible velocities for the target and the scatterers 
yields the total number of collisions. Dividing by $\Delta t$ gives the average
rate.

If the volume occupied by the molecules themselves is an appreciable
fraction of the container volume, then it is incorrect to assume that the 
local density in the collision cylinder is $n$. The inadequacy of $n$ is
explained by the observation that, given a sphere, it is impossible to find
another sphere at radial separation less than a molecular diameter, whereas
for slightly larger separation there tends to be an increased likelihood of
finding a sphere relative to the case where an equivalent volume is examined at
random in the gas. While there is no attractive force between the spheres,
this surplus in density is produced because two nearby spheres tend to receive
collisions on all sides except those sides facing each other resulting in an
external pressure that tends to keep the molecules from separating. Such an
effect leads to an effective attraction between the spheres.\cite{hockey}
Since the collision frequency is proportional to the number density in the
immediate neighborhood of a target molecule we must replace $n$ by $n\chi$,
where $\chi > 1$ represents the ratio of the local density of molecular centers
at the point of contact to the average density throughout the container. 
In a uniform gas $\chi$ depends only on the reduced density $n(2a)^3$.

The above discussion implies that
\begin{equation}
\omega = \sigma n\chi \iint d\mathbf{v}_1\, d\mathbf{v}_2\,
|\mathbf{v}_1-\mathbf{v}_2| f(\mathbf{v}_1) f(\mathbf{v}_2).
\end{equation}
A change of variables from 
$(\mathbf{v}_1, \mathbf{v}_2)$ to
$(\mathbf{v}_1, \mathbf{v}_\textrm{rel} \equiv \mathbf{v}_1 - \mathbf{v}_2)$,
the absolute value of the Jacobian being unity,
and the evenness of the velocity distribution allows us to write
\begin{equation}
\omega = \sigma n\chi \int d\mathbf{v}_\textrm{rel}\, v_\textrm{rel} 
\int d\mathbf{v}_1\, f(\mathbf{v}_1)f(\mathbf{v}_\textrm{rel} - \mathbf{v}_1).
\end{equation}
Since the inner integral defines the equilibrium 
probability density for the relative velocity, then the outer integral
simply computes the mean relative speed in the gas. Therefore,
\begin{equation}
\lambda_\textrm{Maxwell} \equiv \frac{\ev{v}}{\omega}
= \frac{\ev{v}}{\ev{v_\textrm{rel}}'}\frac{1}{\sigma n\chi},
\end{equation}
where $\ev{\dotsc}'$ denotes the average in the relative velocity
distribution.\cite{relative}
A straightforward calculation shows that the ratio 
$\ev{v}/\ev{v_\textrm{rel}}' = 1/\sqrt{2}$.

Although we computed the average collision frequency $\omega$ using a time
average there is a clever argument given by Einwohner and Alder that shows
how to formulate this calculation in terms of an ensemble
average.\cite{Einwohner} Their derivation is described in the appendix.
The meaning of $\chi$ is very clear in this approach: $\chi$ is the radial
distribution function for pairs of spheres evaluated just outside their point
of contact.

The average collision frequency $\omega$ is also a statistical mean,
in the equilibrium velocity distribution, of the collision rate $r(v)$
which describes the probability per unit time that a given molecule with
speed $v$ encounters another molecule. We assume, by virtue of the
low density of gas, that this collision rate is independent of the past history
of the molecule. In other words, the probability that a given molecule suffers
a collision between an arbitrary time $t$ and $t + dt$ is $r(v)dt$. So
\begin{equation}
\label{meancollrate}
\omega \equiv \ev{r(v)} = \int d\mathbf{v}\, f(v) r(v),
\end{equation}
and this exposes Maxwell's mean free path as the ratio of two averages:
\begin{equation}
\label{Maxwell}
\lambda_\textrm{Maxwell} = \frac{\ev{v}}{\ev{r(v)}}.
\end{equation}
It is not obvious that this ratio is the first moment of the probability
distribution for free path lengths.
Historically, alternative definitions for the mean free path have been
suggested. Tait's 
definition takes the distance that a particle of a preselected speed travels
from a given instant in time until its next collision and averages over the
equilibrium distribution of velocities,\cite{Jeans}
\begin{equation}
\label{Tait}
\lambda_\textrm{Tait} \equiv \ev{\frac{v}{r(v)}}.
\end{equation}
Another definition, which is possible from dimensional considerations, is to
first average the inverse collision rate over all
velocities and multiply by the mean speed,
\begin{equation}
\label{other}
\lambda_\textrm{other} \equiv \ev{v}\ev{\frac{1}{r(v)}}.
\end{equation}

The explicit formula for $r(v)$ is known and it is detailed in the classic
texts of the subject.\cite{Jeans,Chapman}
Although this formula is not needed to understand the central claim in this
article, in Section \ref{collratesphere} we present a self-contained
derivation of this rate for hard spheres in three dimensions that serves as an
alternative to the traditional approach of scattering theory. A particularly
lucid discussion of the probability rate $r(v)$ from the traditional viewpoint
may be found in Ref.~\onlinecite{Reif}. Using the
explicit formula for the speed-dependent collision rate one finds 
that definitions~(\ref{Tait}) and
(\ref{other}) differ from Maxwell's by about 4\%.\cite{Masius}
See Table \ref{mfps}.

\begin{table}[h!]
\centering
\caption{Various mean free paths}
\begin{ruledtabular}
\begin{tabular}{l c}
due to & $\lambda (\sigma n\chi)$ \\
\hline
Maxwell & 0.7071 \\
Tait & 0.6775 \\
other & 0.7340 \\
\end{tabular}
\end{ruledtabular}
\label{mfps}
\end{table}

Our primary goal is to emphasize a point which has not been stressed in the
literature: Eq.~(\ref{Maxwell}) may be regarded as the mean of the
probability distribution for lengths of free paths in a hard-sphere gas.
The problem of constructing such a distribution was thoroughly analyzed by
Visco, van Wijland, and Trizac\cite{ViscoShort, ViscoLong} building
upon a key earlier observation by Lue.\cite{Lue}
In addition to studying the distribution analytically, Visco et al.
used computer simulations to study a hard-disk gas in two dimensions.
They obtained excellent agreement between their analytical and numerical
results. We shall review their construction and see how it leads, almost
trivially, to the claim about Maxwell's mean free path.

In Section II we construct the probability distribution for the free path
time and in Section III we generalize that procedure to find 
the distribution for free path length.
In Section IV we obtain the collision rate for hard spheres
and in Section V we use that rate to discuss explicit results for the
hard-sphere gas.

\section{Probability distribution for the time of a free path}

The first order of business
is to find the probability $P(t)dt$ that any given molecule, after
surviving a time $t$ since its last collision, suffers a collision in the
interval $t$ to $t + dt$. To this end 
select a collision at random from the gas and focus
attention on one of the molecules involved in that collision. Following 
Refs.~\onlinecite{ViscoShort,ViscoLong}, we may call this the
``tagged'' molecule.

Prior to the collision the tagged molecule has maintained
some constant velocity $\mathbf{v}$ for a time $t$. The conditional probability
that a molecule with velocity $\mathbf{v}$
has survived for at least time $t$ is $p(t|v) = \exp(-r(v)t)$.\cite{survive}
By multiplying
this by $r(v)dt$ one obtains the probability
that the molecule will suffer a collision during an infinitesimal time interval
succeeding time $t$:
\begin{equation}
\label{survival}
p(t|v) \; r(v)dt.
\end{equation}

It remains to characterize how likely different velocities $\mathbf{v}$
are for the tagged molecule. Since the tagged molecule is obtained by first
identifying a collision, the probability distribution must account for the
likelihood of finding a velocity in the range $\mathbf{v}$ to $\mathbf{v} + 
d\mathbf{v}$ in \textit{any randomly selected collision}. That probability, 
$f_\textrm{coll}(v)d\mathbf{v}$, may be found as follows. At any given moment
there is a fraction $f(v_1)d\mathbf{v}_1$ of the gas that has velocity in the
range $\mathbf{v}_1$ to $\mathbf{v}_1+d\mathbf{v}_1$, and of that a fraction
$r(v_1)dt$ will undergo a collision in the next short time interval $dt$.
Therefore, the relative likelihood of the tagged molecule having some
velocity centered around $\mathbf{v}_1$ as opposed to, say, some $\mathbf{v}_2$,
is given by the ratio
\begin{equation}
\frac{f_\textrm{coll}(v_1)d\mathbf{v}_1}{f_\textrm{coll}(v_2)d\mathbf{v}_2}
= \frac{r(v_1)dt f(v_1)d\mathbf{v}_1}{r(v_2)dt f(v_2)d\mathbf{v}_2}.
\end{equation}
This implies that the tagged molecule's velocity is obtained from the 
probability distribution
\begin{equation}
\label{oncoll}
f_\textrm{coll}(v)d\mathbf{v} \equiv \omega^{-1} r(v) f(v)d\mathbf{v},
\end{equation}
where the factor $\omega^{-1}$ is required by normalization and the definition
given by Eq.~(\ref{meancollrate}). Lue refers to Eq.~(\ref{oncoll})
as the ``on-collision'' velocity distribution.
Molecules with larger-than-typical speeds
collide more often than those with smaller-than-typical speeds, so in any
collision one is more likely to find molecules with larger velocities than 
suggested by the nominal equilibrium velocity distribution.
The distribution is thus a product
of a Gaussian and a weight that enhances the relative probability of finding
faster particles.\cite{effuse} This skewing of the nominal distribution can be
seen explicitly --- in Section \ref{results} it is shown that for a 
hard-sphere gas the on-collision speed distribution has
a most probable value slightly greater than that for the Maxwell speed
distribution.

Sampling velocities from the equilibrium velocity
distribution would be inconsistent
with the requirement that the time $t$ represents the entire time of free
flight of the tagged molecule since that entails choosing a particle at random
at any moment in its flight rather
than some moment a very short time preceding an encounter with another
sphere. In the former case the time $t$ would exclude the remaining time
left on the particle's linear path to its next collision.

The desired probability distribution for the time $t$ of a free path is
obtained by multiplying the probability to find a molecule with velocity
centered on $\mathbf{v}$ that has just undergone a collision at time zero,
with the probability that it will survive for some time $t$ without receiving a
collision, with the probability that it will suffer its next collision
between time $t$ and $t + dt$, and finally summing over all possible velocities.
This amounts to multiplying expressions (\ref{oncoll}) and (\ref{survival})
and integrating over $\mathbf{v}$:
\begin{equation}
P(t)dt = \int f_\textrm{coll}(v)d\mathbf{v} \; p(t|v) \; r(v)dt.
\end{equation}
Visco et al. point out that this distribution is consistent with the 
sensible requirement that the mean time between subsequent collisions is equal
to the reciprocal of the mean collision rate defined in
Eq.~(\ref{meancollrate}). That is, $\int_0^\infty \, tP(t)dt = \omega^{-1}$.

\section{Probability distribution for the length of a free path}

By a similar construction we may obtain the probability distribution 
$P(\ell)d\ell$ for the length $\ell$ of a free path.
$P(\ell)d\ell$ is formed from a product of three terms:
the probability to find a molecule with velocity centered on $\mathbf{v}$ in a
collision selected at random at time zero, the probability that it will
survive without collision for at least a time equal to $t = \ell/v$, and the
subsequent probability to suffer a collision between $t = \ell/v$ and $t = 
(\ell + d\ell)/v$. One must then sum over all possible velocities. Therefore,
\begin{eqnarray}
  P(\ell)d\ell
  & = &
  \int f_\textrm{coll}(v)d\mathbf{v} \;
  p\bigl(t=\ell/v\bigl|v\bigr) \;
  r(v)\frac{d\ell}{v} \nonumber \\
  \implies
  P(\ell) & = & 
  \frac{1}{\omega}\int d\mathbf{v} \frac{r(v)^2}{v}f(v)e^{-r(v)\ell/v}.
\label{pathlengthdist}
\end{eqnarray}
The expectation value of $\ell$ in this distribution is easily evaluated by
exchanging the order of integration: the $\ell$-integral evaluates to
$(v/r(v))^2$ and the remaining integral defines the mean speed in the Gaussian
distribution,
\begin{equation}
\int_0^\infty \ell P(\ell) d\ell = \frac{\ev{v}}{\omega}.
\end{equation}
This is identical to Eq.~(\ref{Maxwell}).

\section{Derivation of collision rate for hard-spheres}
\label{collratesphere}

The hard-sphere gas is a particularly clean system because collisions are
instantaneous and it is unnecessary to consider
simultaneous collisions of three or more spheres. We consider a gas that has
attained steady state and is spatially uniform for which the H-theorem implies
that the velocity distribution $f(\mathbf{v})$ is independent of $t$ and
$\mathbf{r}$. That is, $f(\mathbf{v})$ is not changed by molecular collisions.
Let our gas consist of $N$ molecules 
in a box of volume $L^3$ at temperature $T$.
Imagine that a given molecule, labeled 1, has traveled some distance $\ell$
between collisions. It has been moving with constant velocity $\mathbf{v}_1$ so
that the time spent traveling in a straight line is $t = \ell/|\mathbf{v}_1|$.
During that time all $N-1$ other molecules have passed by without making
contact. Using reasoning based on excluded volume we will calculate the
probability that no other molecule hits 1 during this time. In this framework
it is also natural to include the lowest order finite-density correction in
$n \equiv N/L^3$. That correction may be found by following
the procedure outlined in the classic text by Chapman and Cowling.\cite{Chapman}

\begin{figure}[h!]
\centering
\includegraphics{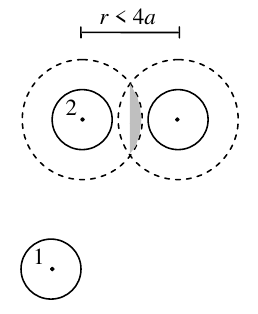}
\caption{The ``shielding'' effect. If a third molecule is sufficiently close to
molecule 2, then the surface area of molecule 2's associated sphere (dashed)
which is accessible to molecule 1's center is reduced by a dome (grayed). For
further discussion of this effect see Ref.~\onlinecite{hockey}.}
\label{shield}
\end{figure}

At the moment
two molecules collide the distance between their centers is $2a$. 
Around each molecule we may draw a concentric ``associated sphere'' with radius
$2a$. A collision occurs when the center of a molecule lies on the 
associated sphere of another molecule. Moreover, that center can never lie
within another's associated sphere. The associated sphere construction
is helpful in two respects: \textit{(i)} it shows
that the inhabitable volume for the center of
molecule 1 is $L^3$ less the total volume taken up by the $N-1$ other associated
spheres;\cite{overlap}
\textit{(ii)} two molecules with overlapping associated spheres shield their
common area in such a way that no other molecule's center can lie on that area.
See Fig.~\ref{shield}.

Since 
we can only be certain that molecule 1 has maintained a constant velocity during
time $t$, it makes sense to use the rest frame of molecule 1 in which to
evaluate the probability that other molecules miss molecule 1. It will be 
convenient to imagine that the center of molecule 1 occupies a single point,
say, the origin of our box. During time $t$ the associated sphere of molecule 2
sweeps out a piecewise cylindrical volume equal to 
$\pi(2a)^2|\mathbf{v}_2 - \mathbf{v}_1|t$.\cite{endcap}
The probability that the origin is not in the
swept-volume of 2's associated sphere would appear to be
\begin{equation}
\label{prob2miss}
p_2 =
\frac{\bigl[L^3 - \frac{4}{3}\pi(2a)^3(N-1)\bigr] 
- \sigma|\mathbf{v}_2 - \mathbf{v}_1|t}{L^3 - \frac{4}{3}\pi(2a)^3(N-1)}.
\end{equation}

However, due to the previously mentioned shielding effect
there is a reduced likelihood of molecule 2
colliding with molecule 1 since, at any given moment, only a fraction of the 
surface area of molecule 2's associated sphere is available to make contact with
molecule 1's center. 
That fraction is explicitly computed in Ref.~\onlinecite{Chapman} and
what follows is an exposition of their procedure. The probable number
of molecular centers within a range $r$ to $r + dr$ of the center of any chosen
molecule is $n 4\pi r^2 dr$.\cite{pair}
Note that for a spatially uniform gas this number
is independent of the positions of other molecules. Two overlapping associated
spheres whose centers are separated by a distance $2a < r < 4a$ obstruct, on
either sphere, a dome with lateral area $4\pi a (2a - r/2)$.
Therefore, the probable area unavailable to a molecule is
\begin{equation}
\int_{2a}^{4a} n 4\pi r^2 dr \cdot 4\pi a (2a - r/2)
= \frac{11}{3}\pi^2(2a)^5 n.
\end{equation}
The available fractional area of molecule 2's associated sphere which can
collide with molecule 1's center is thus
\begin{equation}
\label{available}
\frac{4\pi(2a)^2 - \frac{11}{3}\pi^2(2a)^5 n}{4\pi(2a)^2}
= 1 - \frac{11}{12}\pi(2a)^3n.
\end{equation}

Expression~(\ref{available}) must multiply the swept-volume of molecule 2's
associated sphere. As a check, if expression~(\ref{available}) is zero then
there is unit probability that the origin lies outside the swept-volume.
The correct modification of expression~(\ref{prob2miss}) is therefore
\begin{eqnarray}
p_2 & = & 
\frac{\bigl[L^3 - \frac{4}{3}\pi(2a)^3(N-1)\bigr]
- \bigl[\sigma|\mathbf{v}_2 - \mathbf{v}_1|t\bigr]
\bigl[1 - \frac{11}{12}\pi(2a)^3n\bigr]}{L^3 - \frac{4}{3}\pi(2a)^3(N-1)}
\nonumber \\
& = & 1 - \frac{\sigma n\chi|\mathbf{v}_2-\mathbf{v}_1|\ell/|\mathbf{v}_1|}{N},
\end{eqnarray}
where
\begin{equation}
\chi
\equiv \frac{1-\frac{11}{12}\pi(2a)^3n}{1-\frac{4}{3}\pi(2a)^3(n-L^{-3})}
= 1 + \frac{5}{12}\pi(2a)^3n + o(n^2) + o(L^{-3}).
\end{equation}
In a large-volume limit the correction of $o(L^{-3})$ is subleading and may be
neglected. Moreover, we have neglected corrections imposed by the boundary.
We see that effect \textit{(i)} tends to increase the probability of collisions
whereas effect \textit{(ii)} tends to decrease it. Higher order corrections
to $\chi$ are known.\cite{Chapman}

A similar argument can be made concerning the probability $p_3$ that the
swept-volume of molecule 3's associated sphere does not contain the origin.
We assume the probabilities $p_2$, $p_3$, etc.
to be independent, essentially invoking the molecular chaos assumption.
However this does not preclude the $N-1$ particles from 
interacting with each other during time $t$; their collisions
will inevitably ``kink'' each other's cylinders. 
The probability that none of the $N-1$ associated spheres hits molecule 1's
center during this time is given by the product of the individual probabilities:
\begin{equation}
p(\ell|\mathbf{v}_1) = \prod_{i=2}^N p_i = 
\prod_{i=2}^N 
\biggl[1 - \frac{\sigma n\chi|\mathbf{v}_i-\mathbf{v}_1|/|\mathbf{v}_1|}{N}\ell
\biggr].
\end{equation}
In this product the values of $\mathbf{v}_i$ are obtained from the equilibrium
velocity distribution. One way to estimate this product is to discretize the 
velocity distribution into a finite number of bins and assign to each 
$\mathbf{v}_i$ a particular value representative of a bin. In the $j$th
bin there will be $Nf(\mathbf{v}_j)d^3v_j$ particles. Thus,
\begin{eqnarray}
p(\ell|\mathbf{v}_1) 
& = & 
\prod_{\text{bins}~j} \Biggl[1 - 
\frac{\sigma n\chi|\mathbf{v}_j-\mathbf{v}_1|/|\mathbf{v}_1|}{N}\ell
\Biggr]^{N f(\mathbf{v}_j)d^3v_j} \nonumber \\
& = &
\exp \sum_j -\sigma n\chi \ell
\frac{|\mathbf{v}_j-\mathbf{v}_1|}{|\mathbf{v}_1|} f(\mathbf{v}_j)d^3v_j 
\nonumber \\
& = & 
\exp\biggl[-\frac{\sigma n\chi \ell}{|\mathbf{v}_1|}\int d^3v\, f(\mathbf{v})
|\mathbf{v}-\mathbf{v}_1|\biggr]. \label{proballmiss}
\end{eqnarray}
In the second step we took the limit of large $N$ (and large volume so as
to keep the concentration fixed and small compared to 1)
and in the third step the sum was generalized to a continuous integral over all
possible velocities. The collision rate is found by comparing 
Eq.~(\ref{proballmiss}) with
$p(t|\mathbf{v}_1) = \exp(-r(v_1)\ell/|\mathbf{v}_1|)$.
We obtain the well-known formula
\begin{equation}
\label{rateprecursor}
r(v_1) = \sigma n\chi \int d^3v\, f(\mathbf{v}) |\mathbf{v}-\mathbf{v}_1|.
\end{equation}
As one might expect the probability of collision is high when the cross
section is large, the gas is dense, or if the relative speed is
large.\cite{Reifsrate}
The integral can be done by orienting the $z$-axis along $\mathbf{v}_1$ so that
the magnitude of the relative velocity $\mathbf{v}-\mathbf{v}_1$ may be computed
from the law of cosines as $\sqrt{v^2 + v_1^2 - 2vv_1\cos\theta}$, where
$\theta$ is the polar angle. Doing the angular and radial integrations in order
produces
\begin{eqnarray}
r(v_1) & = &
\sigma n\chi
\sqrt{\frac{2}{\pi}}\int_0^\infty dv\, v^2 e^{-\half v^2}
\times
\begin{cases}
v + \frac{v_1^2}{3v} & v > v_1\, , \\
v_1 + \frac{v^2}{3v_1} & v < v_1\, ,
\end{cases} \nonumber \\
& = &
\sigma n\chi \Biggl[
\sqrt{\frac{2}{\pi}}\exp\biggl(-\frac{v_1^2}{2}\biggr) + 
\bigl(v_1 + v_1^{-1}\Bigr)\text{erf}\biggl(\frac{v_1}{\sqrt{2}}\biggr)\Biggr].
\label{rate}
\end{eqnarray}
Here $\text{erf}$ is the error function.
The collision rate is finite as $v_1 \to 0$, monotone increasing, and
behaves asymptotically as $r(v_1)/\sigma n\chi \sim v_1 + o(1/v_1)$. The fact
that $r(v_1)$ scales as $v_1$ for large speed is completely expected based on
the fact that the majority of molecules have relatively low speeds compared to
the very few molecules in the tail of the Maxwell distribution. For $v_1$ much
larger than the characteristic speed set by the temperature it is
as if all molecules except for molecule 1 are frozen in place. Then it is
easy to see that the frequency of scattering events incurred by molecule 1 is
directly proportional to its speed. Notice that the term in Eq.~(\ref{rate})
which dominates the asymptotic behavior does indeed come from the $v_1 > v$
domain of integration. More to the point, if we approximate the relative speed
$|\mathbf{v}-\mathbf{v}_1|$ by just $v_1$ in Eq.~(\ref{rateprecursor}), then
the integral evaluates to $v_1$.

\section{Explicit results for the hard-sphere gas}
\label{results}

\begin{figure}[h!]
\centering
\includegraphics{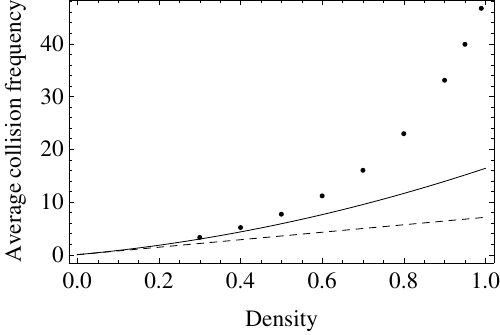}
\caption{Average collision frequency in the hard-sphere gas versus density.
The frequency is given in units of $(kT/m)^{1/2}(2a)^{-1}$ and the density is 
measured in units of $(2a)^{-3}$. 
The solid line is the predicted rate $\omega$ from Eq.~(\ref{meanrate}) using
$\chi = 1 + \frac{5}{12}\pi (2a)^3n$, and the dashed line is the predicted rate
with $\chi = 1$. The dots are the reciprocal of the average time between
collisions taken from Table I of Ref.~\onlinecite{Lue}.
The density range shown is between 0, the extreme dilute limit, and 1, which
is somewhat 
close to the limit of close packing of the spheres within the container.
The fractional error between the predicted and simulated collision frequency
grows with increasing density. It is less than 10\% only for $n(2a)^3 < 0.3$.
}
\label{ratesim}
\end{figure}

The collision rate found in Eq.~(\ref{rate}) may be averaged in the equilibrium
velocity distribution to obtain the mean collision rate:
\begin{equation}
\label{meanrate}
\omega = \ev{r(v)} = \frac{4}{\sqrt{\pi}}\sigma n\chi.
\end{equation}
We may compare predictions for this rate with the numerical results of
molecular dynamics simulations of a hard-sphere gas over a wide range
of density. In Fig.~\ref{ratesim} we plot Eq.~(\ref{meanrate}) against data
from Ref.~\onlinecite{Lue}. As expected, Eq.~(\ref{meanrate}) deviates from
the actual collision frequency when the density of the gas increases. 
The discrepancy is less than 10\% for densities below $0.3(2a)^{-3}$.

\begin{figure}[h!]
\centering
\includegraphics{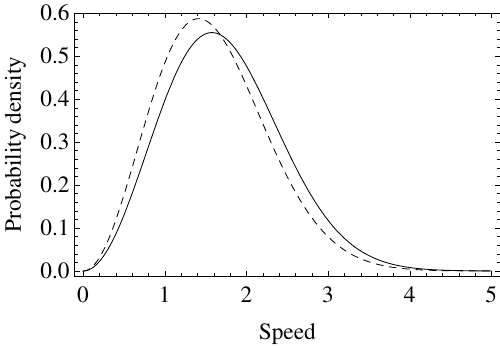}
\caption{Different types of probability density functions for speeds.
Speed is measured in units of $(kT/m)^{1/2}$ and each density function has
the inverse unit. The dashed curve represents the Maxwellian probability
density $4\pi v^2 f(v)$ for randomly picking a molecule with speed $v$ from the
gas. The solid curve represents the probability density
$4\pi v^2 f_\textrm{coll}(v)$ for finding a molecule with incoming speed $v$
when one member of a colliding pair is examined at random.
This figure appears in Ref.~\onlinecite{Lue}.}
\label{collpdf}
\end{figure}

The on-collision distribution in terms of
speed, $f_\textrm{coll}(v)4\pi v^2 dv$, is evidently independent of the
molecular size and density of the gas. The mean speed is
$(3+\pi)/(2\sqrt{\pi}) \approx 1.7325$ and the most probable speed is found
numerically to be $1.5769$. These
may be contrasted with the corresponding values $\sqrt{8/\pi} \approx 1.5958$
and $\sqrt{2} \approx 1.4142$, respectively, for the Maxwell speed distribution,
which are lower. The on-collision speed distribution 
results from multiplying the Maxwell speed distribution
by the monotonically increasing function $r(v)/\omega$ which skews the
distribution toward larger $v$. These two distributions are plotted in
Fig.~\ref{collpdf}.

The free path length distribution given in Eq.~(\ref{pathlengthdist}), when
appropriately normalized, may be expressed in terms of a scaling function.
That is,
\begin{equation}
\lambda_\textrm{Maxwell} P(\ell/\lambda_\textrm{Maxwell})
= F(\ell/\lambda_\textrm{Maxwell}),
\end{equation}
where
\begin{equation}
F(x) \equiv \frac{1}{\pi}\int_0^\infty \frac{dv}{v} \psi(v)^2
\exp\biggl[-v^2 - \frac{1}{\sqrt{2\pi}} v^{-2} \psi(v) x\biggr],~~~~
\psi(v) \equiv v e^{-v^2} + (2v^2 + 1)\int_0^v e^{-t^2} dt.
\label{F}
\end{equation}
The scaling function $F$ does not depend on any dimensionless parameters other
than its argument which is the ratio of the free path length and the mean free
path. In particular, $F$ is independent of the reduced density $n(2a)^3$.
This striking density independence of the scaled free path length distribution 
was noticed in early molecular dynamics simulations of fluids.\cite{Alderletter}

\begin{figure}[h!]
\centering
\includegraphics{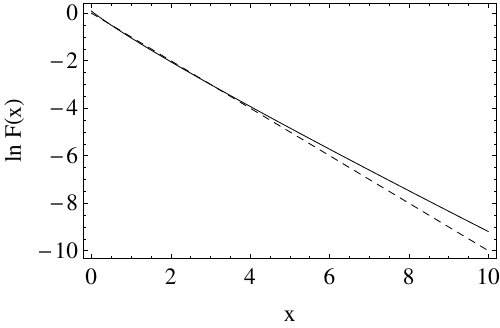}
\caption{Logarithm of the scaling function for the free path length density.
The solid curve is $\ln F(x)$. In the text it is shown that the distribution
of free path lengths, $P(\ell)d\ell$, when parametrized in units of the mean
free path as $x = \ell/\lambda_\textrm{Maxwell}$, is given by $F(x)dx$.
For contrast the dashed curve is a straight line with negative unit slope. 
The disagreement between $\ln F(x)$ and $-x$ shows that the free path length
density is not a simple decaying exponential. Its behavior simplifies somewhat
for asymptotically large $x$: the dominant term is exponential and of the form
$\ln F(x) \sim -x/\sqrt{2}$, although this slope is only apparent for
$x \sim o(100)$ and higher.\cite{ViscoShort, ViscoLong}
}
\label{scaling}
\end{figure}

A plot of $\ln F$ is shown in Fig.~\ref{scaling} and it is contrasted with a
plot of $-x$ in order to show the deviations from pure exponential behavior.
The special result $F(x) = e^{-x}$ would only result if all molecules in the
gas were stationary except for the one molecule executing a free path. This
would correspond to a collision rate $r(v) \propto v$ for which the computation
of Eq.~(\ref{pathlengthdist}) can be done explicitly and yields a simple
exponential. To evaluate the long-distance behavior of $F(x)$ given by 
Eq.~(\ref{F}) one can use the method of steepest descent.\cite{steepest}
It turns out that $F(x) \sim e^{-x/\sqrt{2}}$.
Further discussion of the distribution may be found in the original
Refs.~\onlinecite{ViscoShort, ViscoLong}.

\begin{figure}[h!]
\centering
\includegraphics{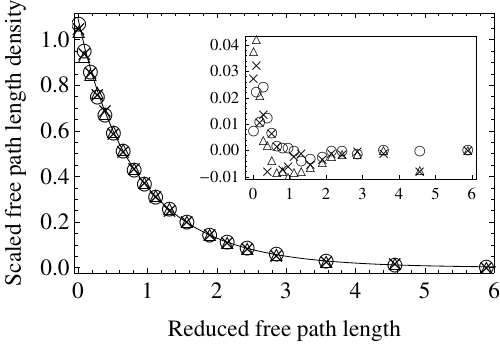}
\caption{Scaling for the distribution of free path lengths.
The solid curve is $F(x)$. In the text it is shown that the distribution
of free path lengths, $P(\ell)d\ell$, when parametrized in units of the mean
free path as $x = \ell/\lambda_\textrm{Maxwell}$, is given by $F(x)dx$. The
data is obtained from Table VIII of Ref.~\onlinecite{Einwohner}. The numerical
simulations were done at three densities: $n(2a)^3 \approx 0.47$ (triangles),
$0.71$ (crosses), and $0.88$ (circles). This figure appears in
Ref.~\onlinecite{Alderletter} but with a subtle difference.\cite{mistake}
The inset shows the difference between the theoretical curve and simulated data.
}
\label{scaleddata}
\end{figure}

In Fig.~\ref{scaleddata} we plot the scaling function $F$ and data obtained from
Einwohner and Alder's early molecular dynamics simulation of the hard-sphere
fluid at three different densities. Even for very dense fluids
the agreement is excellent.\cite{mistake}

\appendix*

\section{The meaning of $\chi$}

In Appendix A of Ref.~\onlinecite{Einwohner}
Einwohner and Alder prove that the mean collision rate per molecule
$\omega(n)$ in a hard-sphere fluid with number density $n$ 
is $\omega(n) = \omega(n \to 0) \chi(n)$. Here
$\omega(n \to 0) = \frac{4}{\sqrt{\pi}}\sigma n\sqrt{\frac{kT}{m}}$
is the collision frequency in the limit of infinite dilution and
$\chi(n)$ is directly related to the probability that, when given a sphere,
there will be any other one located at a center-to-center distance of $2a$. 
We discuss the salient points of their proof.

Suppose that the $N$ spheres experience a total of $M$ collisions during a 
time $t_0$ which is much longer than the average time between collisions.
The total collision rate may be written as a time average,
\begin{equation}
  N\omega = M/t_0 = \frac{1}{t_0}\int_0^{t_0} dt\, \sum_{k=1}^M \delta(t-t_k),
\end{equation}
where $\{t_k\}$ are the times at which the collisions occur. Imagine that one
knows the trajectories $\mathbf{r}_i(t)$ of all $N$ spheres parametrized by the
time $t$. Then the sum of delta functions in time may be reexpressed as a 
sum of delta functions involving the trajectories,
\begin{equation}
  \sum_{k=1}^M \delta(t-t_k) = \sum_{i=2}^N \sum_{j=1}^{i-1}
  \delta(\mathbf{r}_{ij}(t)^2 - 4a^2)
  \theta(-\mathbf{r}_{ij}\cdot\mathbf{v}_{ij})
  |2\mathbf{r}_{ij}\cdot\mathbf{v}_{ij}|.
\end{equation}
We use the notation $\mathbf{r}_{ij} \equiv \mathbf{r}_i - \mathbf{r}_j$
and a similar one for the relative velocity. The expression in terms of the
particles' phase space coordinates is more complicated since a collision
requires that two conditions be met: that the sphere centers are one diameter
apart 
and that the relative velocity of the spheres must make an obtuse angle with the
separation vector, the vector pointing from the center of sphere $j$ to the
center of sphere $i$. The latter condition is necessary to pick out only those
events in phase space where spheres are just about to collide and to avoid
those events where spheres depart or pass parallel without touching;
it has been incorporated using the unit step function $\theta$ which equals 1
for positive argument and 0 otherwise. Note that a Jacobian factor involving
the relative velocity is required by the change of variables. The equality
of time and ensemble averages is now invoked:
\begin{equation}
  \frac{1}{t_0}\int_0^{t_0} O(t)dt = 
  \frac{\int e^{-H/kT} O(\{\mathbf{r}_i(t)\},\{\mathbf{v}_i(t)\})\,
  d\mathbf{r}_1 \dotsb d\mathbf{r}_N \, d\mathbf{v}_1 \dotsb d\mathbf{v}_N}
  {\int e^{-H/kT}
  d\mathbf{r}_1 \dotsb d\mathbf{r}_N \, d\mathbf{v}_1 \dotsb d\mathbf{v}_N},
\end{equation}
where $H = \half m \sum_{i=1}^N \mathbf{v}_i^2
+ \Phi(\mathbf{r}_1, \dotsc, \mathbf{r}_N)$ is the classical energy of the
system. The potential energy is a sum of two-body terms that only depends on
the separations between particles:
$\Phi = \sum_{i < j} \phi(|\mathbf{r}_{ij}|)$, where $\phi$ is infinite if
$|\mathbf{r}_{ij}|$ is strictly less than $2a$ and zero otherwise. Since the
Boltzmann factor factorizes into a product of kinetic and potential terms
the velocity integrals can be done exactly. The delta function 
then collapses two of the position integrals. One obtains
\begin{equation}
  \omega = 
  \frac{4}{\sqrt{\pi}}\sqrt{\frac{kT}{m}} \,
  \pi(2a)^2 \, 
  \frac{N-1}{V} \,
  \underbrace{
  \frac{V^2 \int d\mathbf{r}_3\dotsb d\mathbf{r}_N\,
  e^{-\Phi(|\mathbf{r}_{12}|=2a, |\mathbf{r}_{13}|, \dotsc)/kT}}
  {\int d\mathbf{r}_1\dotsb d\mathbf{r}_N\,
  e^{-\Phi(|\mathbf{r}_{12}|, |\mathbf{r}_{13}|, \dotsc)/kT}}}_{\chi}.
\end{equation}
For a uniform fluid $\chi$ is a function of the reduced density $n(2a)^3$. 
If we assume that the molecular chaos approximation holds even when the gas is
dense so that the density of molecular centers
in the immediate neighborhood of a molecule is not
correlated to the velocity of that molecule, then we would also expect that the
collision rate $r(v)$ experienced by any molecule moving with speed $v$ is
directly proportional to the same factor $\chi$ derived above. See the
discussion in Ref.~\onlinecite{Chapman}.
In particular, $\chi$ does not depend on $v$.

\begin{acknowledgments}

We gratefully acknowledge helpful discussions with Jacob Morris and
correspondence from Fr\'ed\'eric van Wijland. We thank Laurence Yaffe and both
referees of this journal for their many comments and questions that have
significantly improved and clarified this work.

\end{acknowledgments}


\begin{thebibliography}{5}
\bibitem{cp} The densest packing of spheres results in a periodic array of
spheres that occupy approximately 74\% of the total available volume. This
corresponds to a number density of $n(2a)^3 = \sqrt{2}$.
\bibitem{survive} This is a proper probability, not a density. Note that
$p(0|v) = 1$. It may be obtained by imagining that the finite time $t$ is
discretized into $N$ equal intervals $dt$. Then the probability to not undergo
a collision in $N$ steps is $p(t|v) = (1-r(v)dt)^N = (1-r(v)t/N)^N$.
In the limit that $N \to \infty$, $p(t|v)$ becomes exponential.
\bibitem{effuse} A similar reasoning is used to obtain the molecular speed
distribution of a hot gas effusing through a small hole in an oven. Recall
that the Maxwell speed distribution is weighted by an additional factor of
the speed reflecting the fact that faster molecules bombard the hole with 
greater frequency than slower molecules --- that relative frequency is the 
ratio of their speeds.
\bibitem{overlap} This is an approximation based on the observation that the
number of overlapping associated spheres in a dilute gas is small compared
to the total number of spheres. In order to correct for the possibility of
overlap we note that the probable number of molecules close enough to any
given molecule for their associated spheres to overlap is proportional to $n$.
Therefore, the probable excluded volume on any molecule scales as $n$.
And there are order $N$ molecules. Thus, the overlap correction to the naive
inhabitable volume estimate appears at $o(n^2)$ in the factor $\chi$.
\bibitem{endcap} It is unnecessary to include contributions to the volume
from the hemispherical endcaps. This is accounted for when we reduce the amount
of inhabitable volume of molecule 1 due to the volumes of the other molecules.
\bibitem{pair} Actually, the right expression is $n g(r)4\pi r^2dr$, where 
$g(r)$ is the radial distribution function for the hard-sphere gas. Expanding
in the density, $g(r) = 1 + o(n)$. The lowest order term, 1, is sufficient to
calculate $\chi$ to $o(n)$ accuracy.
\bibitem{Reifsrate} In Ref.~\onlinecite{Reif} a remark given on p.~470 
summarizes a rigorous traditional derivation of the collision rate $r(v_1)$
that is more general than the one given in this article. Reif's expression 
involves an additional integration over all solid angle of a differential
scattering cross section which can be computed for an arbitrary
interaction potential between molecules.
\bibitem{steepest} For very 
large $x$ the main contribution to the integral comes
from an environment of $v$ wherein the function $-(2\pi)^{-1/2}v^{-2}\psi(v)$,
which multiplies the factor $x$ in the exponent of the integrand, goes through
a maximum. This happens as $v \to \infty$.
Therefore, the dominant asymptotic behavior of the integral is
of the form $\exp(bx)$, where
$b = \lim_{v \to \infty} -(2\pi)^{-1/2}v^{-2}\psi(v) = -1/\sqrt{2}$.
\bibitem{mistake} In Ref.~\onlinecite{Einwohner} the authors
actually formulate the scaled free path length distribution without the
additional factor of $r(v)/\omega$ since they incorrectly assumed that the
probability to find a molecule with a given speed in a collision was given
by the Maxwell speed distribution. In our notation they found
$F_\textrm{incorrect}(x) = \frac{2\sqrt{2}}{\pi}\int_0^\infty dv\, \psi(v)
\exp\bigl[-v^2 - \frac{1}{\sqrt{2\pi}} v^{-2} \psi(v) x\bigr]$.
Somewhat fortuitously, the error incurred by using the wrong distribution 
makes a very small adjustment to the shape of the scaling function.
The absolute value of the difference between $F$ and $F_\textrm{incorrect}$
is no greater than about 0.07 and this occurs for small values of the
argument. Such discrepancy is noticeable in Fig. 1 of
Ref.~\onlinecite{Alderletter}. Note that the correct scaling 
function predicts $F(0) \approx 1.0921$ which matches the data better at very
small free path length than is reported in that letter.
\end{thebibliography}
\end{document}